\documentclass[a4paper,english,useAMS,usenatbib]{mn2e}
\usepackage[T1]{fontenc}
\usepackage[latin9,utf8]{inputenc}
\usepackage[active]{srcltx}
\usepackage{color}
\usepackage{verbatim}
\usepackage{booktabs}
\usepackage{amstext}
\usepackage{graphicx}
\usepackage[authoryear]{natbib}

\makeatletter

\providecommand{\tabularnewline}{\\}

\@ifundefined{date}{}{\date{}}

\newcommand{\uct}{1}
\newcommand{\ox}{2}
\newcommand{\rhod}{3}
\newcommand{\sarao}{4}
\newcommand{\idiauwc}{5}
\newcommand{\uwc}{6}
\newcommand{\cea}{7}
\newcommand{\inafm}{8}
\newcommand{\herts}{9}
\newcommand{\csiro}{10}
\newcommand{\jaA}{11}
\newcommand{\jaB}{12}
\newcommand{\yaA}{13}
\newcommand{\yaB}{14}
\newcommand{\brienz}{15}
\newcommand{\inafb}{16}
\newcommand{\mbA}{17}
\newcommand{\idiauct}{18}
\newcommand{\syd}{19}
\newcommand{\nmA}{20}
\newcommand{\durh}{21}
\newcommand{\saao}{22}
\newcommand{\nrao}{23}
\newcommand{\smrA}{24}
\newcommand{\pret}{25}
\newcommand{\icrar}{26}
\newcommand{\GEPI}{27}
\newcommand{\RATSS}{28}
\newcommand{\USN}{29}

\title[GRGs in MIGHTEE-COSMOS]{\textbf{MIGHTEE: Are giant radio galaxies more common than we thought?}}
\author[J. Delhaize et al.]{J. Delhaize$^{\uct}$\thanks{E-mail:jacinta@ast.uct.ac.za}, {I. Heywood$^{\ox,\rhod,\sarao}$}, {M. Prescott$^{\idiauwc}$}, {M.J.~Jarvis$^{\ox,\uwc}$}, {I. Delvecchio$^{\cea,\inafm}$},  \newauthor{I.H. Whittam$^{\ox,\uwc}$}, S.V. White$^{\rhod}$, M.J. Hardcastle$^{\herts}$, C.L. Hale$^{\csiro}$,
 J. Afonso$^{\jaA,\jaB}$,\newauthor Y. Ao$^{\yaA,\yaB}$, M. Brienza$^{\brienz,\inafb}$, M. Br\"uggen$^{\mbA}$, J.D. Collier$^{\idiauct,\syd}$, E. Daddi$^{\cea}$, M. Glowacki$^{\idiauwc,\uwc}$,\newauthor N. Maddox$^{\nmA}$, L.K. Morabito$^{\durh}$, I. Prandoni$^{\inafb}$, Z. Randriamanakoto$^{\saao}$, S. Sekhar$^{\idiauct,\nrao,\idiauwc}$, \newauthor
Fangxia An$^{\idiauwc}$, N.J. Adams$^{\ox}$, S. Blyth$^{\uct}$, R.A.A. Bowler$^{\ox}$, L. Leeuw$^{\uwc}$, L. Marchetti$^{\uct,\inafb}$,\newauthor S.M. Randriamampandry$^{\saao,\smrA}$,   K. Thorat$^{\pret,\idiauct}$, N. Seymour$^{\icrar}$ O. Smirnov$^{\rhod,\sarao}$,\newauthor A.R. Taylor$^{\idiauct,\idiauwc}$, C. Tasse$^{\GEPI,\RATSS,\USN}$  M. Vaccari$^{\idiauwc,\inafb}$ 
\\Affiliations are listed at the end of the paper}

\makeatother

\usepackage{babel}

\date{Accepted 2020 December 9. Received 2020 November 18; in original form 2020 September 23}
\pubyear{2020}

\begin{document}

\maketitle
\pagerange{\pageref{firstpage}--\pageref{lastpage}} \pubyear{2020}

\label{firstpage}
\begin{abstract}
We report the discovery of two new giant radio galaxies (GRGs) using
the MeerKAT International GHz Tiered Extragalactic Exploration (MIGHTEE)
survey. Both GRGs were found within a $\sim$1$\,$deg$^{2}$ region
inside the COSMOS field. They have redshifts of $z=0.1656$ and
$z=0.3363$ and physical sizes of 2.4$\,$Mpc and 2.0$\,$Mpc, respectively. Only the cores of these GRGs were clearly visible in previous high resolution VLA observations, since the diffuse emission of the lobes was resolved out. However, the excellent sensitivity and {\em uv} coverage of the
new MeerKAT telescope allowed this diffuse emission to be detected. The GRGs occupy an unpopulated region of radio power -- size parameter space. 
Based on a recent estimate of the GRG number density, the probability of finding two or more GRGs with such large sizes at $z<0.4$ in a $\sim$1$\,$deg$^{2}$ field is only $2.7\times10^{-6}$, assuming Poisson statistics. This supports the hypothesis that the prevalence of GRGs has been significantly underestimated in the past due to limited sensitivity to low surface brightness emission. The two GRGs presented here may be the first of a new population to be revealed through surveys like MIGHTEE which provide exquisite sensitivity to diffuse, extended emission.

\end{abstract}

\section{Introduction}

Some classes of active galactic nuclei (AGN) have jets of relativistic plasma/particles emanating from the central region, which produce radio synchrotron emission. In some
cases, these jets propagate to extremely large distances beyond the
host galaxy and into the intergalactic medium (IGM). When the projected
linear size of the jets and lobes exceeds 0.7\,Mpc, such systems are referred
to as giant radio galaxies (GRGs; e.g.\,\citealp{schoenmakers01,lara01}; \citealp{dabhade20})\footnote{Adjusted to the modern cosmology of \citet{planck16} for consistency with the current literature.}. GRGs are the largest individual objects in the Universe. The largest known has a projected linear size\footnote{Based on a redshift of $z=0.3067$, an angular size of 17.4\,arcmin, and adjusted to our chosen
cosmology.} of 4.89\,Mpc \citep{machalski08}, though the majority  of known GRGs are below 2\,Mpc in extent \citep{dabhade20b}.

Several factors have been proposed to explain why the jets of GRGs are able to extend to such large distances. One suggestion is that these systems exist
in low density environments that allow the jets to permeate easily
through the IGM (e.g. \citealp{mack98,malarecki15}). However, $\sim$10\ per cent
of GRGs have now been found to reside in cluster environments \citep{komberg09,dabhade20b,tang20} and \cite{lan20} recently found no difference between the environments of GRGs and that of galaxy control samples.
Another scenario is that the central engines of GRGs may boast particularly
powerful and/or restarted AGN activity, though several studies including \citet{komberg09} and \citet{hardcastle19} found little evidence that GRGs are different
to normal radio galaxies in this respect. 

A prevailing idea is that these objects represent the oldest AGN systems, such that the
jets have had enough time to grow to their large sizes (e.g.\,\citealp{ishwara-chandra99}). However, we may then expect the existence of many more GRGs than are
currently known \citep{komberg09}.

Fewer than 1000 GRGs have been found to date. \citet{dabhade20b} recently compiled a comprehensive catalogue of all 820 GRGs known. At the time of writing, a further six have been reported by \cite{ishwara-chandra20} and five by \citet{tang20}.

The first GRG discovery was made in the 1970s \citep{willis74} and
since then GRGs have primarily been found via wide-field radio continuum
surveys such as the NRAO VLA Sky Survey (NVSS; \citealp{condon98}),
the Faint Images of the Radio Sky at Twenty-Centimeters survey (FIRST;
\citealp{becker95}), the Westerbork Northern Sky Survey (WENSS;
\citealp{rengelink97}) and the Sydney University Molonglo Sky Survey (SUMSS; \citealp{mauch03}).

The highly-extended nature of GRGs and the generally low surface brightness
of their lobes, which fade as they age, make them notoriously difficult to
detect and identify. For example, 162 GRGs have only very recently
been discovered in NVSS data, despite this survey having already existed
for two decades. These were found via the Search and Analysis of Giant
radio galaxies with Associated Nuclei (SAGAN) project, which rigorously
combined newly-developed automated pattern recognition techniques
\citep{proctor16} with careful manual inspection \citep{dabhade17,dabhade20b}. Despite such efforts, and the existence
of many thousands of `normal' sized ($<0.7$\,Mpc) radio galaxies
(RGs), GRGs remain scarce (e.g. \citealp{kaiser97}). 

However, the new generation of deep and wide-field radio surveys, with sensitivity to a range of spatial scales, may provide a much clearer understanding of the number density and physics of such sources. In particular, low frequency surveys with new-generation
instruments like the Low Frequency Array (LOFAR; \citealp{vanhaarlem13}),
the Murchison Widefield Array (MWA; \citealp{tingay13}) and the upgraded
Giant Metre-wave Radio Telescope (uGMRT; \citealp{gupta2017}) \textbf{
} are proving excellent resources for detecting and characterising
GRGs (e.g.~\citealp{hurley-walker15,hardcastle16,clarke17,seymour20,cantwell20}).
This is due, in part, to the increasing brightness towards lower radio
frequencies often displayed by radio galaxies. Indeed, \citet{dabhade20}
recently reported the highest sky density of consistently-sampled
GRGs using Data Release 1 of the LOFAR Two-metre Sky Survey (LoTSS;
\citealp{shimwell19}). They found 239 GRGs over a 424\,deg$^{2}$
region at 120-168\,MHz.

In the GHz regime, the newly-commissioned MeerKAT telescope in South Africa \citep{jonas16}
is proving to be an excellent instrument for GRG studies (e.g. \citealp{cotton20}).
Although it operates at higher frequencies than LOFAR, MWA and uGMRT (though there is some overlap in observing frequency with the latter), MeerKAT has the excellent sensitivity and {\em uv} coverage ideal for
such work, including the simultaneous availability of long and short baselines.

The MeerKAT International GHz Tiered Extragalactic Exploration survey (MIGHTEE; \citealp{jarvis16}) is a galaxy evolution survey underway with MeerKAT. Among its data products will be high-quality radio continuum data over a relatively wide field (20\,deg$^2$). In this paper, we report the discovery
of two newly-identified GRGs in 1\,deg$^{2}$ MIGHTEE Early Science observations
of the COSMOS field. The discovery of these objects within such a small sky area hints at the presence of a `hidden' population of GRGs,
hitherto undetected due to observational limitations. If GRGs prove
to be more common than previously thought, this may alter our understanding
of the AGN duty cycle and the impact of AGN-induced jets on the evolution of galaxies and
the IGM. 

Throughout this paper we assume a $\Lambda$CDM cosmology with
$H_{0}$$=67.8$$\,$km\,s$^{-1}$\,Mpc$^{-1}$, $\Omega_{\Lambda}=0.692$
and $\Omega_{{\rm M}}=0.308$ \citep{planck16}. We assume a \citet{chabrier03} initial mass function, unless otherwise stated.

\section{Radio continuum surveys\label{sec:The-data-1}}

\subsection{MIGHTEE L-band data}

\begin{table*}
\begin{minipage}{170mm}
	\caption{Summary of the MeerKAT MIGHTEE observations used for this study. The COSMOS field centre is J2000 10h00m28.6s +02d12m21s, and the COSMOS\_8 pointing is at J2000 10h00m29.0s +02d33m33.79s. The primary calibrator for all observations was PKS B1934-638, and the secondary was 3C 237. The observations used MeerKAT's $L$-band system, 900 -- 1670 MHz.}
	\centering
	\label{tab:observations}
	\begin{tabular}{lllllll} 
		\hline
		Date       & Block ID   & Field     & Antennas & Track (h) & On-source (h) & Channels \\ \hline \hline
		2018-04-19 & 1524147354 & COSMOS    & 64                 & 8.65      & 6.1          & 4096  \\
		2018-05-06 & 1525613583 & COSMOS    & 62                 & 8.39      & 5.1           & 4096  \\
		2020-04-02 & 1585844155 & COSMOS\_8  & 60                 & 8         & 6.25          & 32768 \\
		2020-04-26 & 1587911796 & COSMOS    & 59                 & 8         & 6.25          & 32768 \\
		\hline
	\end{tabular}
\end{minipage}
\end{table*}

The MIGHTEE survey targets four extragalactic deep fields, namely the European Large Area ISO Survey - South 1 (ELAIS-S1), COSMOS, the XMM-Newton Large Scale Structure (XMM-LSS) field, and the Extended Chandra Deep Field South (E-CDFS), for a total sky area of approximately 20\,deg$^2$. The survey uses total intensity, polarised intensity and spectral line data products to achieve a range of science goals. It has components using MeerKAT's $L$-band (900 -- 1670 MHz) and $S$-band (1.75 -- 3.5 GHz) receivers, and a primary design requirement is to reach the $L$-band classical confusion limit in total intensity at about 2 $\mathrm{\mu}$Jy beam$^{-1}$ (for a resolution of $\sim8$\,arcsec).

The initial release of total intensity continuum within the MIGHTEE consortium included a single pointing in the COSMOS field (J2000 10h00m28.6s +02d12m21s; see Table \ref{tab:observations}).
The two GRGs we present in this paper were discovered in this pointing. Since one of the GRGs was towards the edge of the field, we have also imaged an extra MIGHTEE pointing (COSMOS\_8; J2000 10h00m29.0s +02d33m33.79s) for which this object was closer to the centre of the primary beam.

Full details of the initial MIGHTEE continuum data release and the data processing method\footnote{The calibration and imaging scripts are available online here: https://www.github.com/IanHeywood/oxkat and through the Astrophysics Source Code Library record ascl:2009.003 \citep{oxcat}} will be presented in Heywood et al. (in prep.), however we provide a brief overview here.

\begin{itemize}

\item{The data were converted from their native format into a MeasurementSet format by the South African Radio Astronomy Observatory (SARAO) archive\footnote{https://archive.sarao.ac.za/}, and averaged from their original \textit{L}-band frequency resolution to 1024 channels in the process. Flags generated by the telescope control and monitoring system were applied.}

\item{Basic flagging commands were applied to all fields using {\sc CASA} \citep{mcmullin07}. Frequency ranges containing persistent radio frequency interference (RFI) were flagged on spacings shorter than 600 m. The auto-flagging algorithms {\sc tfcrop} and {\sc rflag} were used on the calibrator fields.}

\item{The standard calibrator PKS B1934-638 was used to derive delay and bandpass solutions using the relevant {\sc CASA} tasks. This was an iterative process, with rounds of autoflagging on residual visibilities in each iteration.}

\item{The gain solutions derived from the primary were applied to the secondary calibrator (3C 237), and an intrinsic spectral model was derived for the latter. Time-dependent complex gains were then derived from the observations of the secondary using this intrinsic model with the {\sc gaincal} task. This process compensates for the effects of the large fractional bandwidth of MeerKAT, coupled with the fact that the data arrive in MeasurementSet format that only has a single spectral window. The flux scale may be biased if this effect is not taken into account.}

\item{All the gain corrections were applied to the target data, which is then flagged using the {\sc tricolour}\footnote{https://github.com/ska-sa/tricolour} package. Removing the edges of the bandpass where the gain sharply rolls off results in 770 MHz of usable bandwidth. A loss of about 50 per cent of the data in this region is typical following RFI removal, although the RFI occupancy is strongly baseline dependent, and most of this loss occurs on spacings shorter than 1 km.}

\item{The target data were imaged using {\sc wsclean} \citep{offringa14}. The data were imaged blindly with 100,000 iterations, and a clean mask was derived from the resulting image, excluding regions below a local threshold of 6$\sigma$ where $\sigma$ is the pixel standard deviation. The data were re-imaged using this mask.}

\item{The multi-frequency clean components from the masked image were used to predict a visibility model for self-calibration using the {\sc casa} {\sc gaincal} task. Phase corrections were derived for every 64 seconds of data, and an amplitude and phase correction was derived for every target scan, with the solutions for the former applied while solving for the latter. The self-calibrated data were re-imaged using {\sc wsclean} and the mask is refined if necessary.}

\item{Direction-dependent corrections were made by imaging the data with {\sc ddfacet} \citep{tasse18}. The resulting model was partitioned into $\sim$10 directions, constrained by the location of off-axis problem sources, and the need to retain suitable flux in the sky model per direction. The {\sc killms} package \citep[e.g.][]{smirnov15} was then used to solve for a complex gain correction for each direction with a time / frequency interval of 5 minutes / 128 channels. Another run of {\sc ddfacet} reimaged the data, applying the directional corrections.}

\item{Finally, the images were primary-beam corrected by dividing them by an azimuthally averaged Stokes I model, evaluated at 1284~MHz using the {\sc eidos} \citep{asad19} 
package.}

\end{itemize}

MIGHTEE continuum data is imaged twice, with a Briggs' robust parameter of 0.0 and -1.2. This is to deliver a higher sensitivity image as well as a higher angular resolution image, the trade-off for which is a loss of sensitivity due to the down-weighting of the many short spacings that MeerKAT's dense core provides. The COSMOS pointing for a total on-source time of 17.45\,h reaches a thermal noise (measured away from the main lobe of the primary beam) of 1.9 $\mu$Jy beam$^{-1}$ in the robust 0.0 image, with an angular resolution of 8.4\,arcsec $\times$ 6.8\,arcsec (position angle -11.2\,deg). The robust -1.2 image reaches 6\,$\mu$Jy beam$^{-1}$  with an angular resolution of 4.8\,arcsec $\times$ 4.0\,arcsec (position angle -12.63\,deg). 

We imaged the 6.25\,h of COSMOS\_8 data with less aggressive weighting (robust = 0.3) in order to increase the sensitivity to the diffuse emission from the lobes. This image has a thermal noise of 2.5\,$\mu$Jy\,beam$^{-1}$ and an angular resolution of 11.6\,arcsec $\times$ 7.4\,arcsec (position angle -12.6\,deg). The higher resolution map made with COSMOS\_8 data was imaged using a robust parameter of -1.2 and has a thermal noise of 8.5\,$\mu$Jy\,beam$^{-1}$ and an angular resolution of 5.0\,arcsec $\times$ 4.0\,arcsec (position angle -15.44\,deg).

Note that in practice the central region of the lower resolution images (both COSMOS and COSMOS\_8) are limited to root mean squared ({\em rms}) `noise' levels of $\sim$3\,$\mu$Jy beam$^{-1}$  by a combination of thermal noise and classical confusion. The lower resolution images are used for the work presented here, unless otherwise stated.

\subsection{VLA-COSMOS 3\,GHz Large Project}

In this paper, we also use data from the
VLA-COSMOS 3\,GHz Large Project (hereafter VLA-3\,GHz). This project
was presented by \citet{smolcic17a} and is a continuum survey
with the Karl G.\,Jansky Very Large Array (VLA; \citealt{perley11}) covering 2.6\,deg$^{2}$ over the full COSMOS
field. This 384\,h survey was centred at 3\,GHz with a 2\,GHz bandwidth
using three sets of 64 pointings in A and C array. It reached an {\em rms} sensitivity of 2.3$\mu$Jy beam$^{-1}$ 
with 0.75\,arcsec resolution. This is similar in sensitivity to
the Early Science MIGHTEE data, accounting for the different central frequencies
and assuming a standard spectral index for radio galaxies ($\alpha=-0.8$;
where $S_{\nu}\propto\nu^{\alpha}$).

While the angular resolution of the VLA-3\,GHz data is superior to
that of MIGHTEE, it has poorer sensitivity to large-scale emission.
The VLA has up to 351 baselines with a minimum baseline length of 36\,m,
while MeerKAT has up to $2,016$ baselines with a minimum length of
29\,m. The larger number of short baselines, combined with the better instantaneous \textit{uv} coverage provided by its configuration, therefore makes MeerKAT
the better instrument for detecting diffuse, extended emission.

\section{GRG properties}

\begin{figure*}
\includegraphics[scale=0.08]{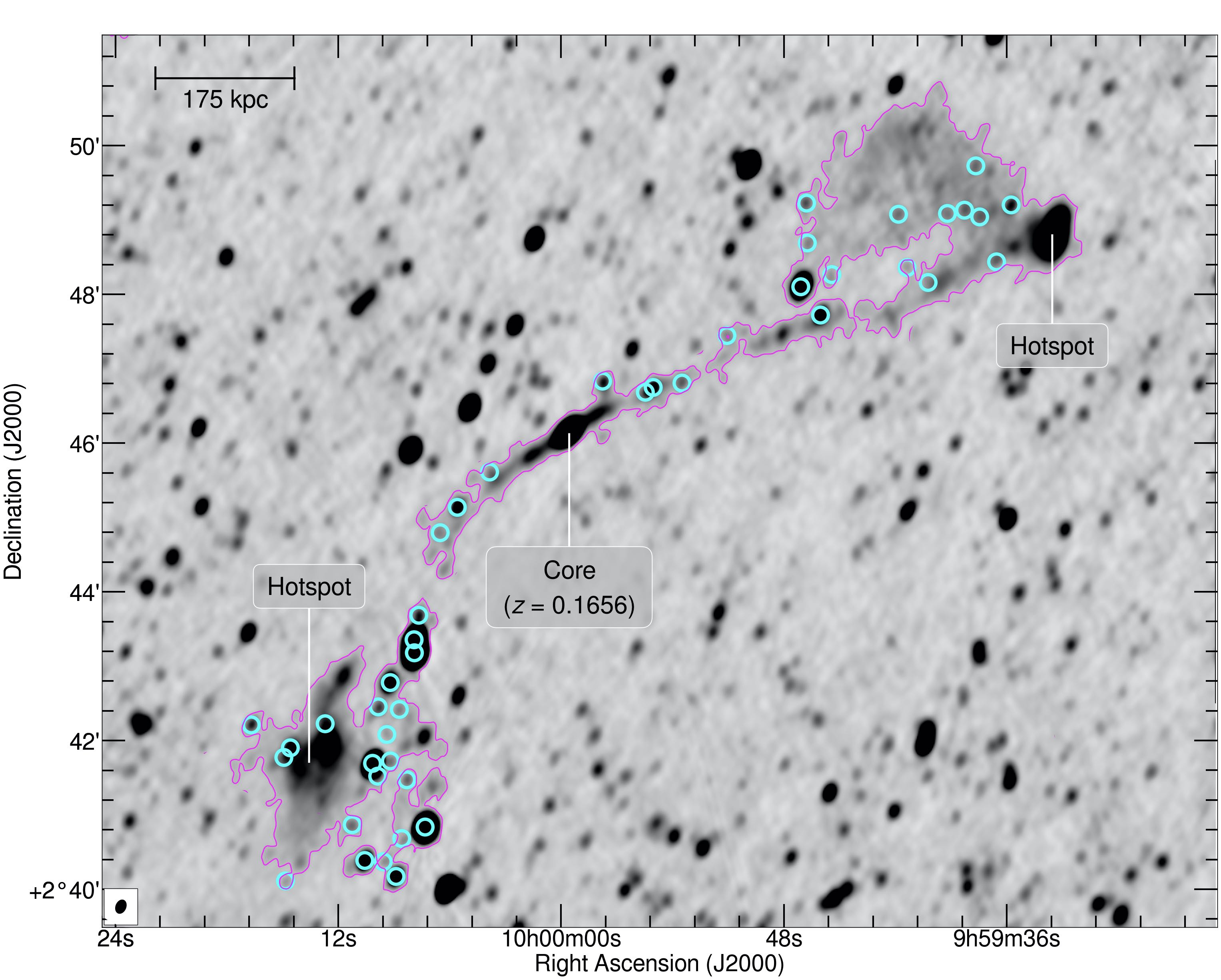}
\includegraphics[scale=0.08]{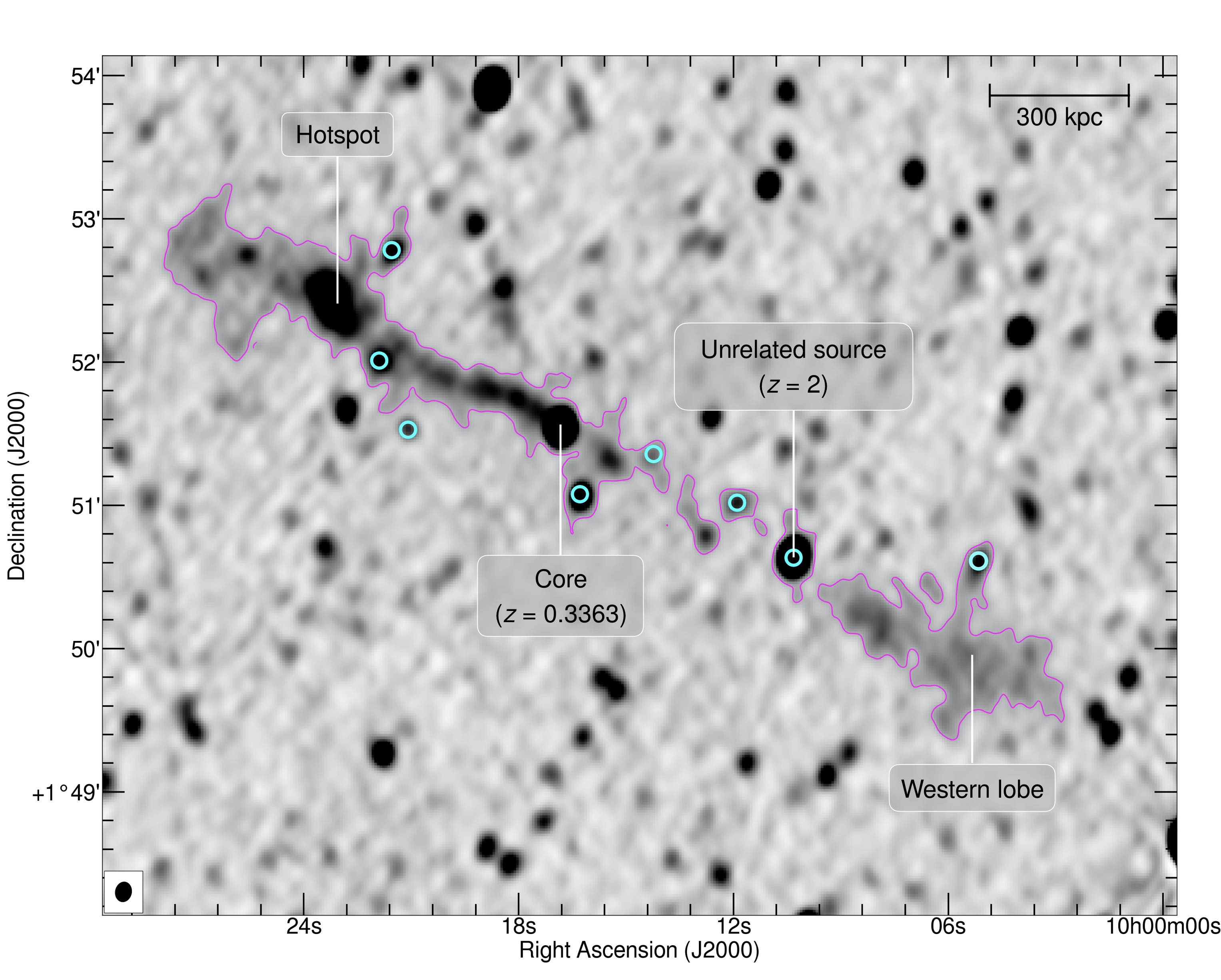}

\caption{GRG1 (left) and GRG2 (right) as seen in MIGHTEE. The magenta contour (for GRG-associated emission only) highlights the full extent of the GRGs. The contour level is $7\,\mu$Jy beam$^{-1}$ for GRG1 and $5\,\mu$Jy beam$^{-1}$ for GRG2. Notable features of the GRGs are labelled. Unrelated continuum sources are marked by cyan circles. The MIGHTEE maps shown are COSMOS\_8 (robust=0.3, thermal noise 2.5$\mu$Jy beam$^{-1}$) for GRG1, and COSMOS (robust=0.0, thermal noise 1.9$\mu$Jy beam$^{-1}$) for GRG2. The beam is shown in the bottom left corner of each image.
\label{fig:grgs-Lband} }
\end{figure*}

\begin{figure*}
\includegraphics[scale=0.15]{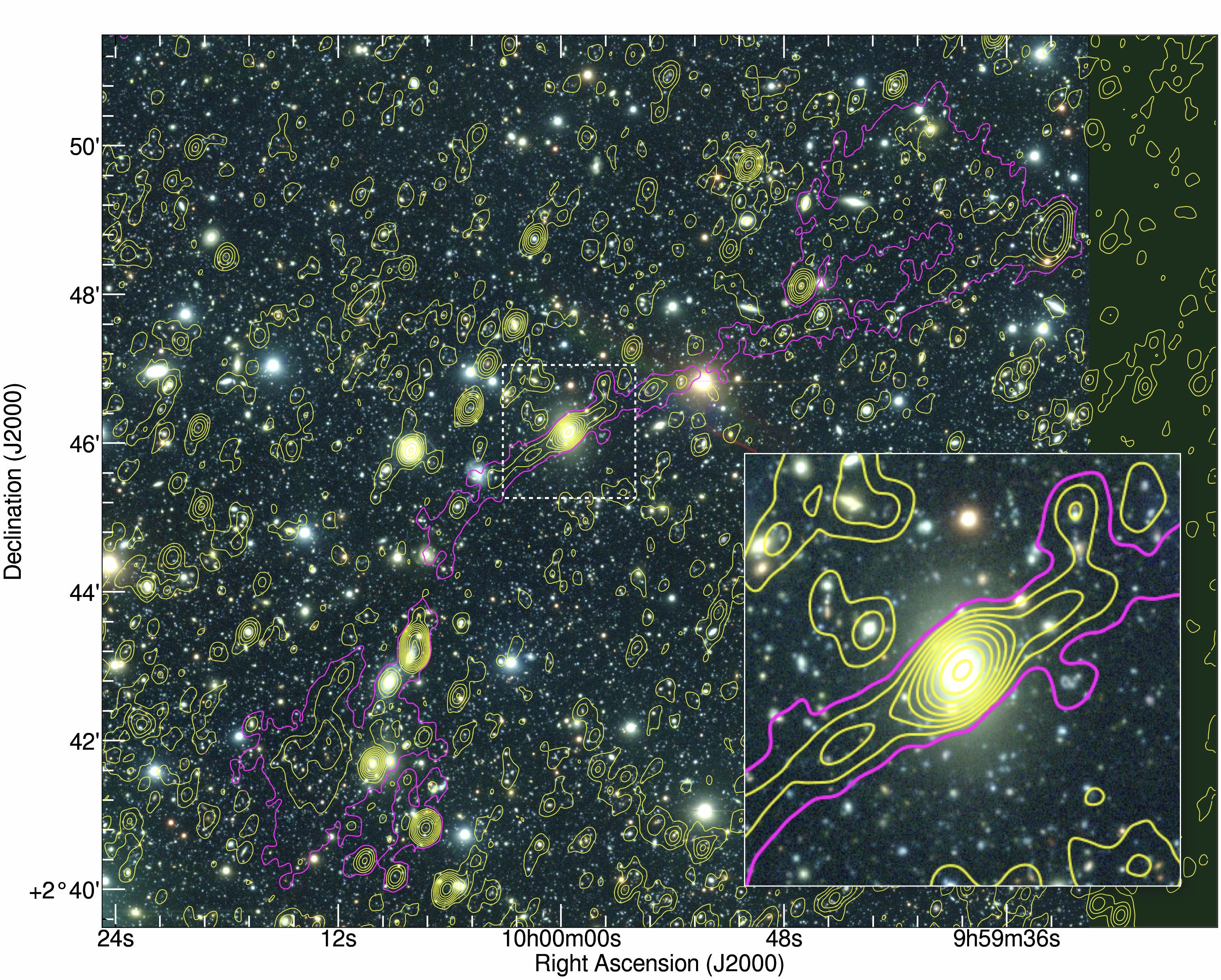}

\caption{GRG1 as seen in MIGHTEE (yellow contours), overlaid on a composite optical image of HSC \emph{g}, \emph{r} and \emph{i} bands. Contours are shown at intervals of  $30\sqrt{3}^n\,\mu$Jy beam$^{-1}$, where $n=0,1,2...$20. The lowest contour level is $7\,\mu$Jy beam$^{-1}$ and is shown in magenta, as in Figure \ref{fig:grgs-Lband}, to highlight the full extent of the GRG. The inset shows an enlargement of the core area, where the host galaxy can be seen.
\label{fig:grg1-optical} }
\end{figure*}

\begin{figure*}
\includegraphics[scale=0.15]{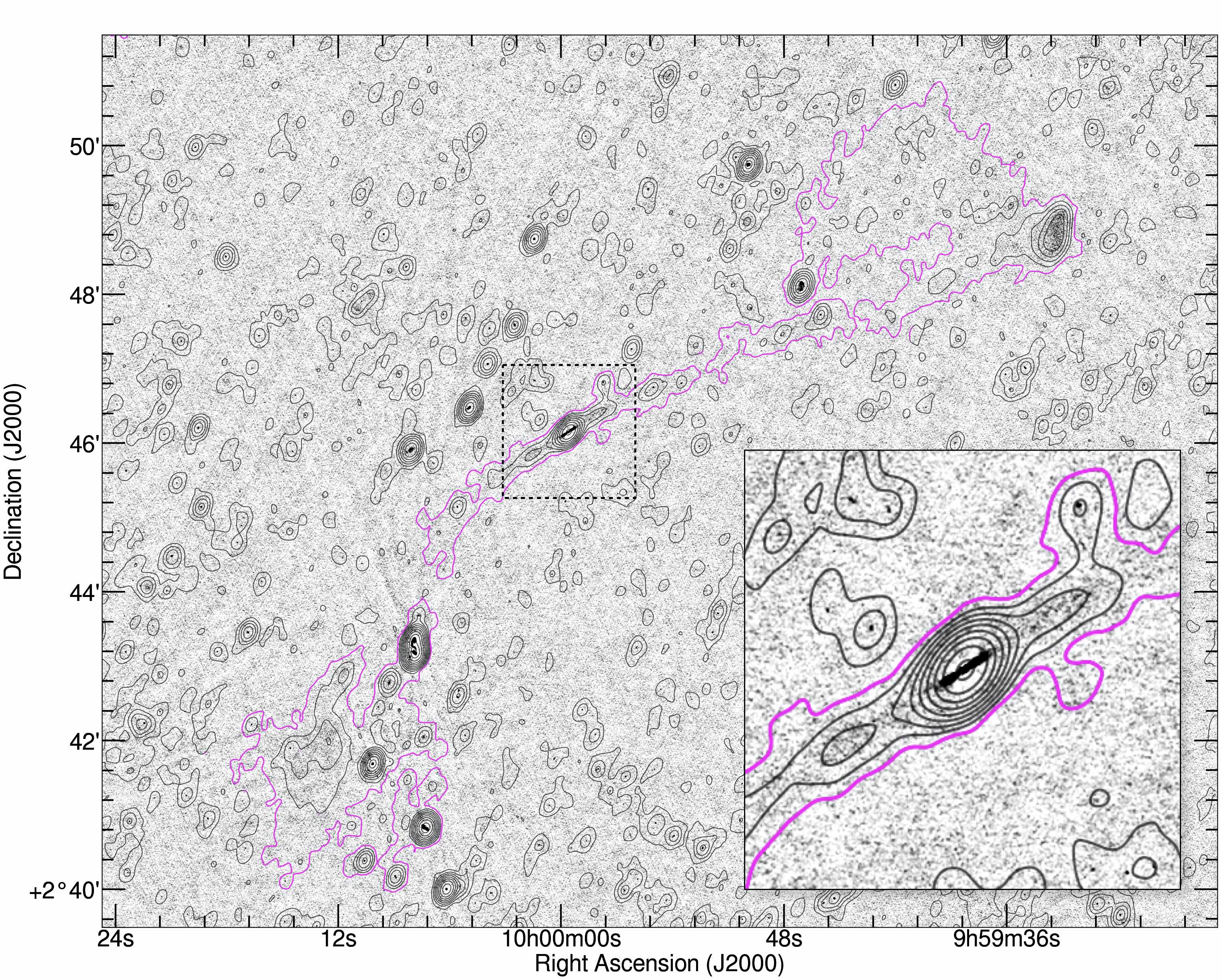}

\caption{GRG1 as seen in the VLA-3$\,$GHz data (background map) and in the MIGHTEE data (black contours). Contour levels are as in Figure \ref{fig:grg1-optical} and the lowest contour level is again shown in magenta. The hot-spot in the northern lobe is seen faintly at 3\,GHz. The inset shows an enlargement of the core region, where a double-lobed/jetted structure can be seen in the VLA-3GHz image.  
\label{fig:grg1-radio}}
\end{figure*}

The two GRGs were serendipitously discovered in the MIGHTEE-COSMOS map (original pointing) during the process of cross-matching the radio sources with their optical counterparts, via visual inspection. Details of this process will be presented in Prescott et al., (in prep), and are similar to those of \citet{prescott18}. Following IAU conventions, we name these sources MGTC\,J095959.63+024608.6 (hereafter GRG1) and  MGTC\,J100016.84+015133.0 (hereafter GRG2). The prefix MGTC indicates the discovery of the sources in the MIGHTEE-continuum survey. The basic properties of each GRG
are summarised in Table \ref{tab:grg-props} and flux densities
and radio powers are presented
in Table \ref{tab:grg-fluxes}. Note that only statistical uncertainties are quoted in Table \ref{tab:grg-props}, but that systematic uncertainties are found to be less than 3 per cent (Heywood et al., in prep). 

\begin{table*}
\caption{Properties of the two GRGs discovered in MIGHTEE-COSMOS. Columns:
(1) Object name (2) Right Ascension (J2000) (3) Declination (J2000) (4) Spectroscopic
redshift (5) Projected angular size (6) Projected linear size. \label{tab:grg-props}}
\begin{centering}
\begin{tabular}{cccccc}
\toprule 
1 & 2 & 3 & 4 & 5 & 6 \tabularnewline
Name & R.A. & Dec & $z$ & $d$ & D\tabularnewline
 &(h:m:s) &(d:m:s) &  & (arcmin) &  (Mpc)\tabularnewline
\midrule
\midrule 
MGTC J095959.63+024608.6 \textit{(GRG1)} & 09:59:59.63 & +02:46:08.6 & 0.1656 & 13.8 & 2.42\tabularnewline
\midrule 
 MGTC J100016.84+015133.0 \textit{(GRG2)} & 10:00:16.84 & +01:51:33.0 & 0.3363 & 6.8 &  2.04\tabularnewline
\bottomrule
\end{tabular}
\par\end{centering}
\end{table*}

\begin{table*}
\caption{Integrated flux and radio power of the two GRGs. Columns:
(1) Object name (2) Effective frequency of MIGHTEE map at position
of GRG (3) Peak brightness of core (4-6) Integrated flux density of core, northern lobe and southern
lobe (7) Total integrated flux density of GRG, combining all components (8) Spectral
index of core (9) Radio power at $\nu_{{\rm eff}}$ ($\sim$1.2\,GHz),
based on columns 7 and 8. $^*$See Table \ref{tab:grg-props} for official names. $^{**}$May include significant levels of unrelated emission from coincident continuum sources. See text for details. \label{tab:grg-fluxes}}
\begin{centering}
\begin{tabular}{ccccccccc}
\toprule 
1 & 2 & 3 & 4 & 5 & 6 & 7 & 8 & 9\tabularnewline
Name$^*$ & $\nu_{{\rm eff}}$ & $S_{{\rm p,core}}$ & $S_{{\rm int,core}} $ & $S_{{\rm int,NL}} $ & $S_{{\rm int,SL}} $ & $S_{{\rm int,all}}$ & $\alpha_{3{\rm GHz}}^{\nu_{{\rm eff}}}$ & $P_{1.2\,\rm{GHz}}$\tabularnewline
 & (GHz) & (mJy\,beam$^{-1}$) & (mJy) & (mJy) & (mJy) & (mJy) &  & (W\,Hz$^{-1}$)\tabularnewline
\midrule
\midrule 
GRG1 & 1.217 & $3.73\pm0.01$ & 5.92$\pm0.01$ & 8.59$\pm0.02$ & $^{**}$21.07$\pm0.01$ & $^{**}$35.59$\pm0.02$ & $-0.51\pm0.03$ & $^{**}(2.438\pm0.002)\times10^{24}$\tabularnewline
\midrule 
GRG2 & 1.256 & $0.72\pm0.01$ & 0.778$\pm0.002$ & 2.04$\pm0.01$ & 0.92$\pm0.01$ & 3.74$\pm0.02$ & $0.29\pm0.04$ & $(1.329\pm0.007)\times10^{24}$\tabularnewline
\bottomrule
\end{tabular}
\par\end{centering}
\end{table*}

\subsection{MGTC J095959.63+024608.6 (GRG1)}

\subsubsection{MIGHTEE data}

GRG1 is shown in Figures \ref{fig:grgs-Lband}, \ref{fig:grg1-optical} and \ref{fig:grg1-radio}.  
The two diffuse outer lobes and jets are detected for the first time in MIGHTEE, thus newly identifying this object as a GRG. The core of GRG1 is located at ${\rm R.A.}={\rm 09h59m59.63s}$ and ${\rm Dec}={\rm +02d46m08.6s}$ and has an elongated structure in MIGHTEE. It is associated with a host galaxy identified in optical and near-infrared data (see Figure \ref{fig:grg1-optical} inset) with a spectroscopic redshift of $z=0.1656$, according to the Sloan Digital Sky Survey Data Release 14 (SDSS DR14; \citealp{abolfathi18}). Summing the distance between the outer edges of each lobe and the central core, the total projected angular size of GRG1 is 13.8\,arcmin. It therefore has a physical size of 2.42\,Mpc, which places it solidly in the regime of giant radio galaxies. 

The northern lobe contains a potential hot-spot which is extended and is not associated with any point source (indicated in Figure \ref{fig:grgs-Lband}). This is confirmed using the higher-resolution (4.8\,arcsec $\times$ 4.0\,arcsec) MIGHTEE image. A comparison with optical data (Prescott et al., in prep) likewise reveals no counterpart. Diffuse
emission extends to the north-east of this hot-spot, in a direction perpendicular to that of the jet. The reason for this is unclear, but could indicate interactions with the surrounding IGM, allowing plasma to flow in that direction (e.g. \citealp{subrahmanyan08}).
This lobe appears to be edge-brightened, which is typical of
a Fanaroff and Riley type II (FRII; \citealt{fanaroff74}) radio galaxy. Some collimated jet emission is present
between the core and the lobes.

There is also evidence of a potential hot-spot towards
the centre of the southern lobe and the presence of a jet. The latter is bent with respect to the northern jet alignment. Again, this could be due to interactions with the
IGM and surrounding environment. For example, \citet{malarecki15}
suggest that the lobes of GRGs could bend to avoid high density regions.

MIGHTEE flux densities in Table \ref{tab:grg-fluxes} are measured by integrating within the region indicated by the magenta contour shown in Figure \ref{fig:grgs-Lband}. The contributions from coincident compact continuum sources not associated with the GRG have been removed. These unrelated sources (presumably fore- or background galaxies) are marked in Figure \ref{fig:grgs-Lband} and were identified in the counterpart cross-matching procedure of Prescott et al., (in prep). This made use of the higher-resolution MIGHTEE and VLA-3GHz data. Except for continuum sources larger than the beam, we assume they are unresolved point sources and approximate their total flux density to be equal to their peak brightness in the higher-resolution MIGHTEE map. This is to minimise the amount of diffuse GRG-related emission removed. 

However, some `contaminating' emission from these unrelated sources may still remain, meaning that the flux density measurements in Table \ref{tab:grg-fluxes} are conservative upper limits, particularly for the southern lobe. A lower limit for the flux density of the southern lobe (avoiding areas with bright continuum sources) is $3.60 \pm 0.01$\,mJy, giving a total flux density for GRG1 of $18.11\pm0.02$\,mJy.

\subsubsection{VLA-3GHz data}

Only the core of GRG1 is clearly detected in the VLA-3GHz data and
the total flux density reported for the object by \citet{smolcic17a} was
$3.02\pm0.15$\,mJy. 

In these high-resolution VLA data, the core of the GRG is resolved
and displays a double-lobed/jetted structure (see inset of Figure \ref{fig:grg1-radio}). These
could be the inner-most part of the jets, or could be a separate set
of inner lobes/jets. In the latter case, this object could be classified
as a double-double radio galaxy (e.g. \citealp{schoenmakers00,mahatma19})
and a candidate restarted AGN (e.g. \citealp{jurlin20, brienza20}
and references therein). However, no hot-spots are evident in this region, as might be expected from inner lobes/jets.

The northern lobe/hot-spot of GRG1 is detected in the 3\,GHz data, as seen in Figure \ref{fig:grg1-radio}.
However, it has a low signal-to-noise ratio (SNR) and therefore does not
appear in the SNR$\geq5$ source catalogue presented
by \citet{smolcic17a}. There is only a vague hint of the southern lobe in
the 3\,GHz map, being highly diffuse and therefore predominantly resolved out of
the VLA data.

\subsubsection{Radio power}\label{GRG1-P}
To determine the spectral index of the core, we first smooth the VLA-3\,GHz map to the resolution of the MIGHTEE map. We then measure the spectral index using the MIGHTEE peak brightness ($3.73\pm0.01$\,mJy\,beam$^{-1}$; to minimise contamination from the jets) and the 3\,GHz peak brightness from the smoothed map ($2.93\pm0.032$\,mJy\,beam$^{-1}$). This gives $\alpha_{3{\rm GHz}}^{\nu_{{\rm eff}}}=-0.51\pm0.03$.
Here, $\nu_{{\rm eff}}=1.217\,$\,GHz and is the effective frequency of the MIGHTEE map at the position of this GRG. The relative flatness of the core's spectrum is expected due to synchrotron self-absorption. 
 
GRG1 has a 1.2\,GHz radio power of $\sim10^{24.4}$\,W\,Hz$^{-1}$ (lower limit of $\sim10^{24.1}$\,W\,Hz$^{-1}$). This was calculated using the measured spectral index of the core and assuming a spectral index of -0.8 for the lobes. This assumption is reasonable for an optically-thin lobe, 
and also given that \citet{dabhade20}
find the mean spectral index distribution of their sample of GRGs
to be $\alpha^{1.4}_{0.144}=-0.79$, which is similar to that of RGs.

Although the morphology of GRG1 somewhat resembles an FRII-like structure, its radio power is more typical of an FRI-type galaxy \citep{fanaroff74}. However, this is consistent with the results of \cite{mingo19} who have recently discovered a population of low-luminosity\footnote{Up to several orders of magnitude below the traditional FR break, which is $\sim 10^{24.5}$\,W\,Hz$^{-1}$ in \emph{L}-band.} FRII-type RGs in LoTSS. They find that radio luminosity does not reliably predict whether a source has an FRI or an FRII-type morphology. Furthermore, GRG1 has a lower radio power than that of most known GRGs. According to \cite{dabhade20b}, known GRGs at $z<1.0$ have mean 1.4\,GHz radio powers of $\sim10^{25.3}$\,W\,Hz$^{-1}$. In many ways, GRG1 is reminicent of NGC6251, a borderline FRI/FRII GRG. Both have weak hot-spots, FRI-like jets and radio powers below the FR break \citep{cantwell20}.

\subsubsection{Host galaxy and AGN characterisation}\label{host-grg1}

\begin{figure*}
\includegraphics[scale=0.7]{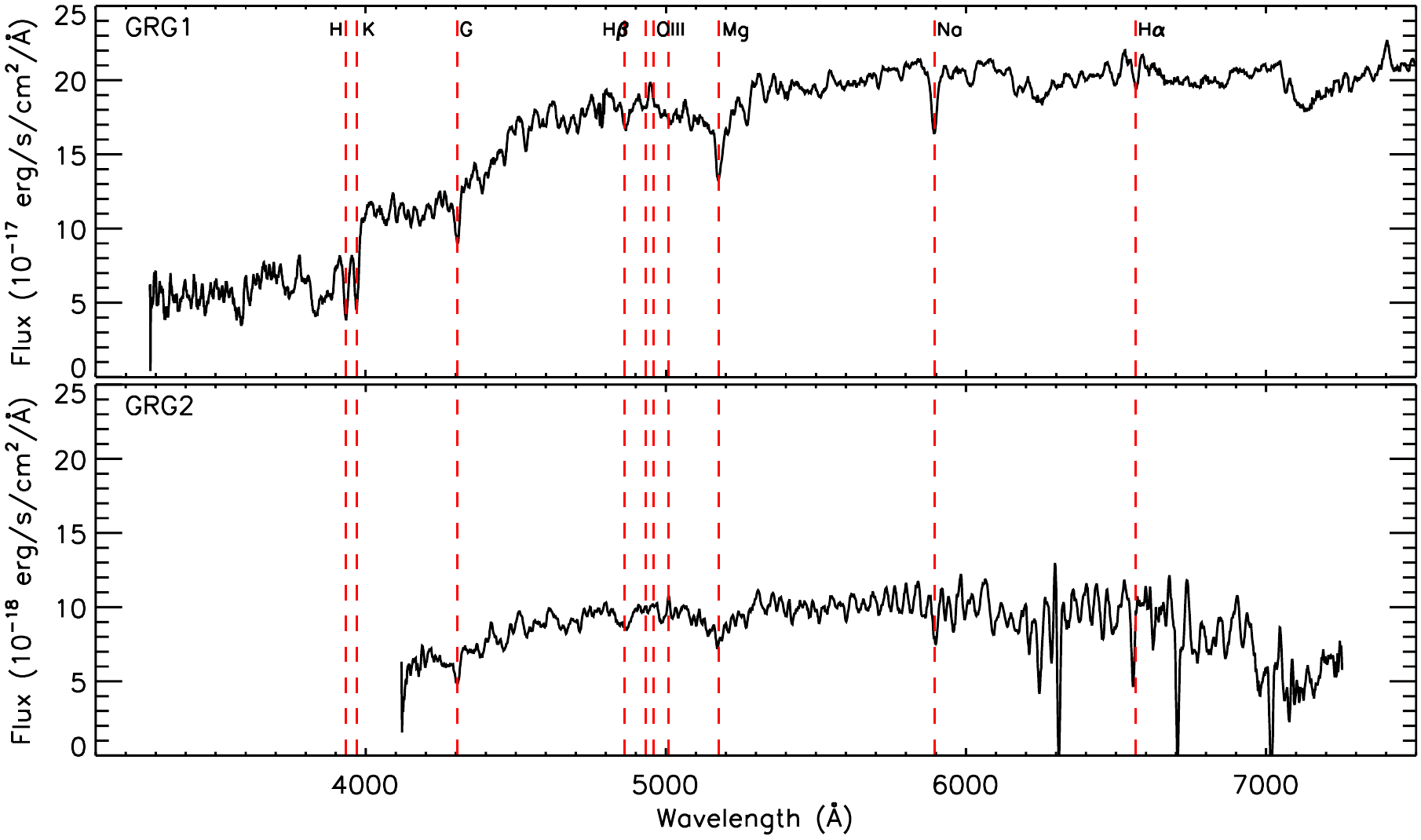}

\caption{Rest-frame optical spectra of the GRG host galaxies. Top: SDSS DR14 spectrum for GRG1. Bottom: zCOSMOS spectrum for GRG2.\textbf{ \label{fig:spectrum}}}

\end{figure*}

The host galaxy of this radio source appears elliptical in Hyper Suprime-Cam imaging (\citealp{aihara19}; see Figure \ref{fig:grg1-optical} inset). The top panel of Figure \ref{fig:spectrum} shows the rest-frame
SDSS DR14 optical spectrum of the host galaxy. Features typical of an early-type elliptical galaxy can be seen in this spectrum, such as a prominent 4000{\AA} break, the presence of strong absorption lines like MgI and NaD, as well as the lack of nebular emission lines. 
It contains no prominent narrow or broad emission lines associated with high excitation radio galaxies or quasars, such as [O{\sc iii}].

To further examine whether there is any evidence of radiatively-efficient AGN activity, we fit the spectral energy distribution (SED) of the host galaxy with various templates. The SED is constructed using the photometric catalogue of \citet{laigle16} (hereafter COSMOS15), along with mid-infrared (including 24$\mu$m) to sub-millimetre data from the "super-deblended" catalogue of \cite{jin18}. Note that the only significant detection (to the 3$\sigma$ level) of GRG1 in {\em Herschel} bands\footnote{Data from the Photoconductor Array Camera and Spectrometer Evolutionary Probe (PEP; \citealp{lutz11}) survey and  {\em Herschel} Multi-tiered Extragalactic Survey (HerMES; \citealp{oliver12}).} is at 100$\mu$m. The 3$\sigma$ upper limits were used in cases of non-detection (see \citealp{delvecchio17} for details).

\begin{figure*}
\includegraphics[scale=0.5]{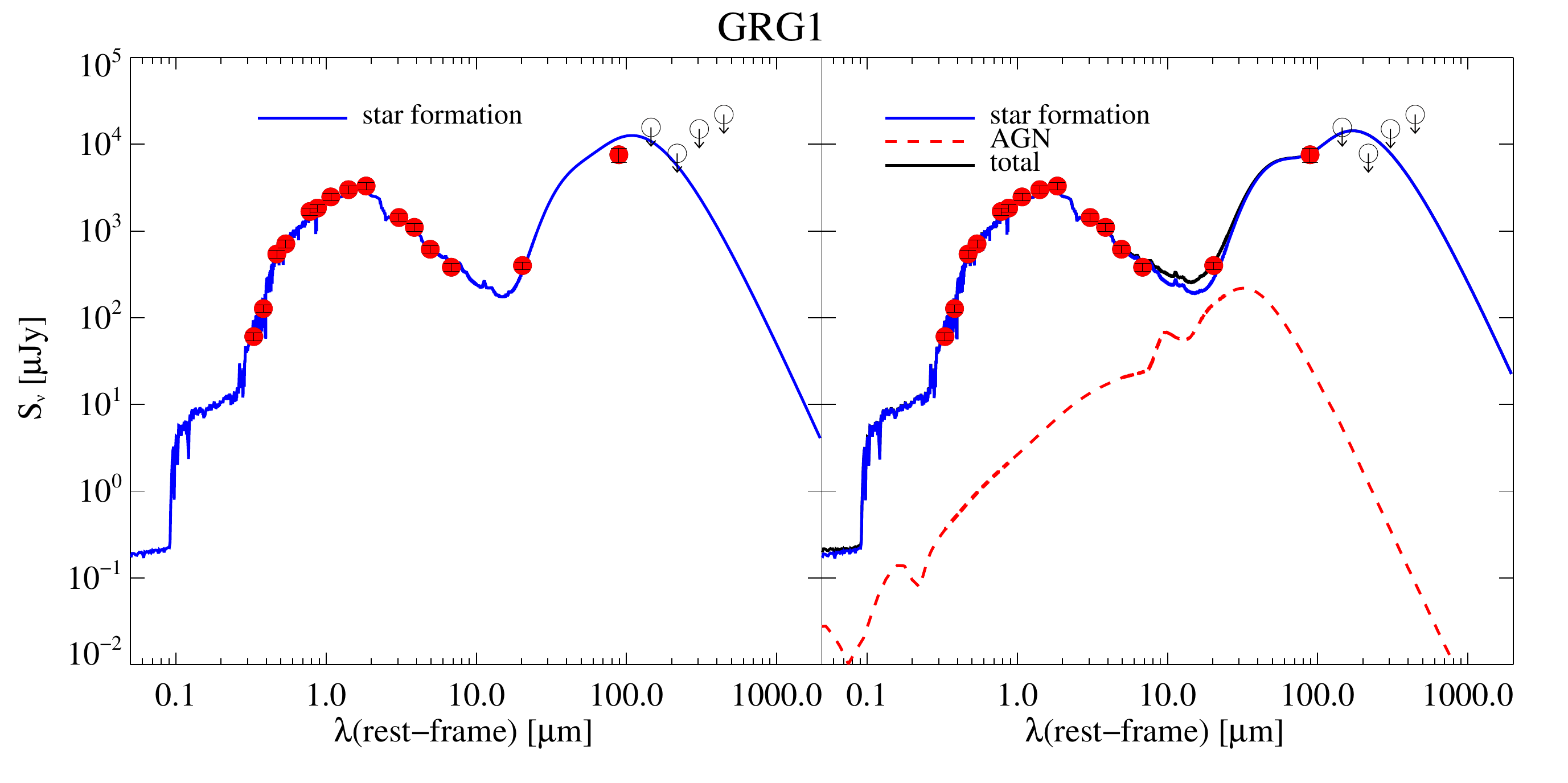}
\includegraphics[scale=0.5]{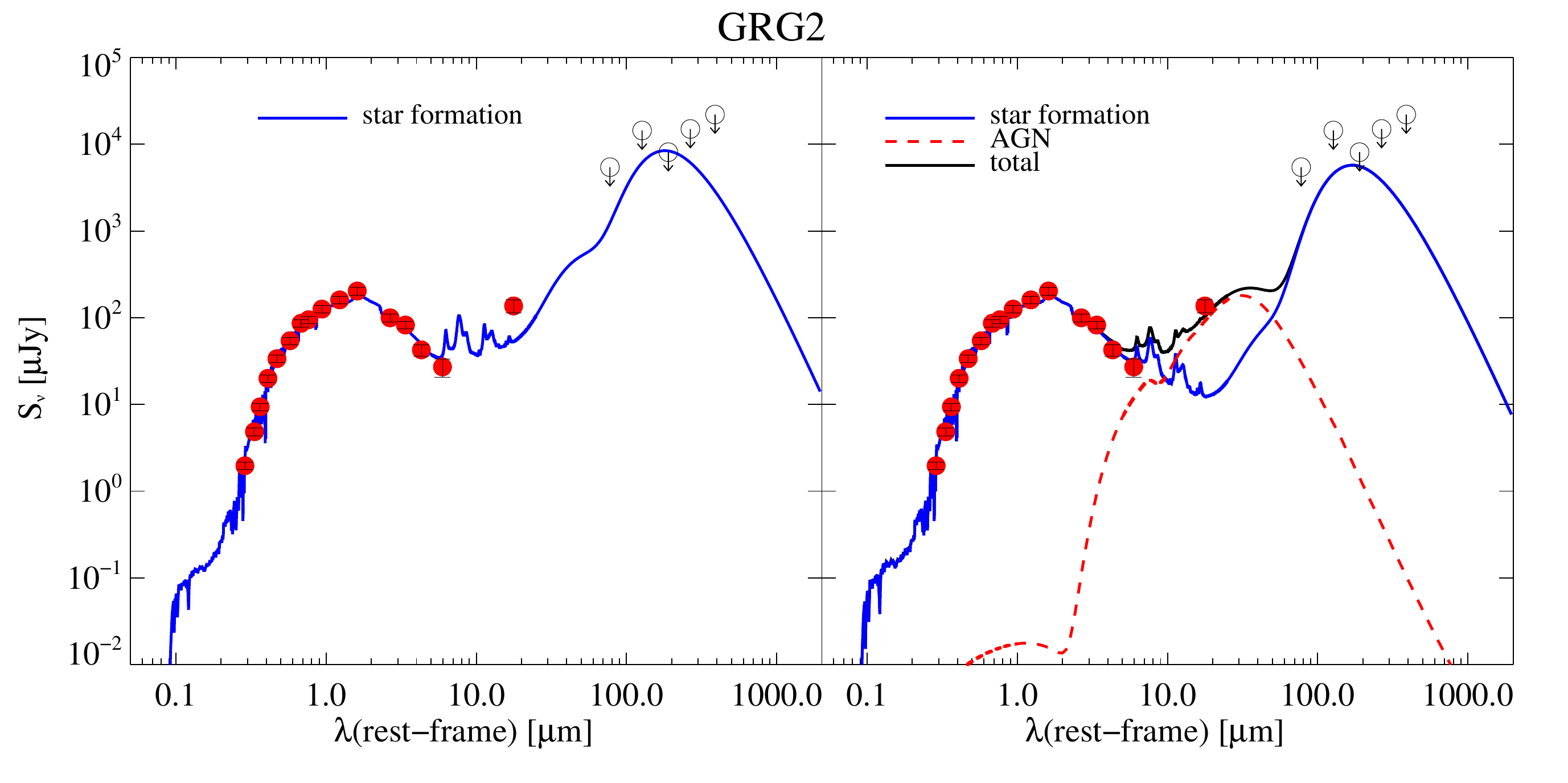}

\caption{SED of the GRG1 (top) and GRG2 (bottom) host galaxies showing detections (closed points) or 3$\sigma$ upper limits (open circles) in UV to sub-millimetre bands. 
The best-fitting template considering only star-formation (blue line) was determined using MAGPHYS (left) and SED3FIT (right). SED3FIT also fits an AGN template (dashed red line), and the combined (AGN + star-forming) template (black line). The MAGPHYS fit on the left was preferred for both GRGs, based on a Fisher-test of the reduced $\chi^2$. See text for further details. \label{fig:SED-of-GRG}}
\end{figure*}

The SED is fit with: (i) the MAGPHYS code of \cite{dacunha08} considering only star-formation, and (ii) the SED3FIT code of \cite{berta2013} which also incorporates a set of AGN templates. The results are shown in Figure \ref{fig:SED-of-GRG} and further details of the SED-fitting approach can be found in \cite{delvecchio17}. The fit obtained in the latter case, with an AGN component, does not significantly (<99 per cent confidence level) improve the reduced $\chi^2$ of the fit, on the basis of a Fisher-test. Therefore, we conclude that SED fitting reveals no evidence of radiatively-efficient AGN activity. 

From the best-fitting parameters of the SED model (without AGN), we find that the host galaxy has a stellar mass of $M_{*}=10^{11.42 }$\,M$_{\odot}$ and a  (combined IR and UV) star formation rate  (SFR) of 1.97\,M$_{\odot}$/yr. It is therefore only weakly star-forming and lies below the main sequence of star formation \citep{schreiber15} by a factor of $\sim$2. This is consistent with the fact that it is undetected in the far-infrared. 

In the X-ray regime, the galaxy has a [2--10] keV luminosity of $L_{X}\sim10^{41.6}$\,erg\,s$^{-1}$, based on data from the Chandra-COSMOS \citep{elvis09,civano12} and COSMOS-Legacy catalogues \citep{civano16,marchesi16}. Since this is roughly five times in excess of that predicted by the SFR-L$_{\rm X-ray}$ relation of star-forming galaxies \citep{lehmer16}, it is possible that concomitant low-luminosity AGN activity is present in the X-ray. The AGN bolometric luminosity predicted by the fit with AGN would be consistent within 1$\sigma$ with the observed $L_{X}$, if assuming a set of [2--10] keV bolometric corrections from \cite{lusso12}. However, the total X-ray emission from the core follows the correlation for unabsorbed jet-related emission in \citet{hardcastle09}. Therefore, this is likely a pure jet system with no evidence for additional accretion-related X-ray emission. 

Therefore, no evidence of (obscured or unobscured) radiatively-efficient AGN activity is found in any regime. Rather, GRG1 can be classified as a low-excitation radio AGN (LERG; e.g. \citealp{best12}) powered by radiatively-inefficient Bondi accretion of hot gas from the IGM (\citealp{hardcastle20} and references therein) hosted by a massive, weakly star-forming elliptical galaxy.

The host of GRG1 is the most massive galaxy in a group of eight according to the zCOSMOS group catalogue of \citet{knobel12} (group ID 606). Furthermore, \citet{giodini10} associated the core of this galaxy with an X-ray cluster. This is unsurprising since GRGs are commonly found to reside in small groups of galaxies \citep{malarecki15}. This environmental information supports the scenario in which the bending of the southern jet, and the extended emission of the northern lobe, result from interactions with the surrounding intergalactic and intragroup/intracluster medium of the GRG. 

\subsection{MGTC J100016.84+015133.0 (GRG2)}

\begin{figure*}
\begin{centering}
\includegraphics[scale=0.15]{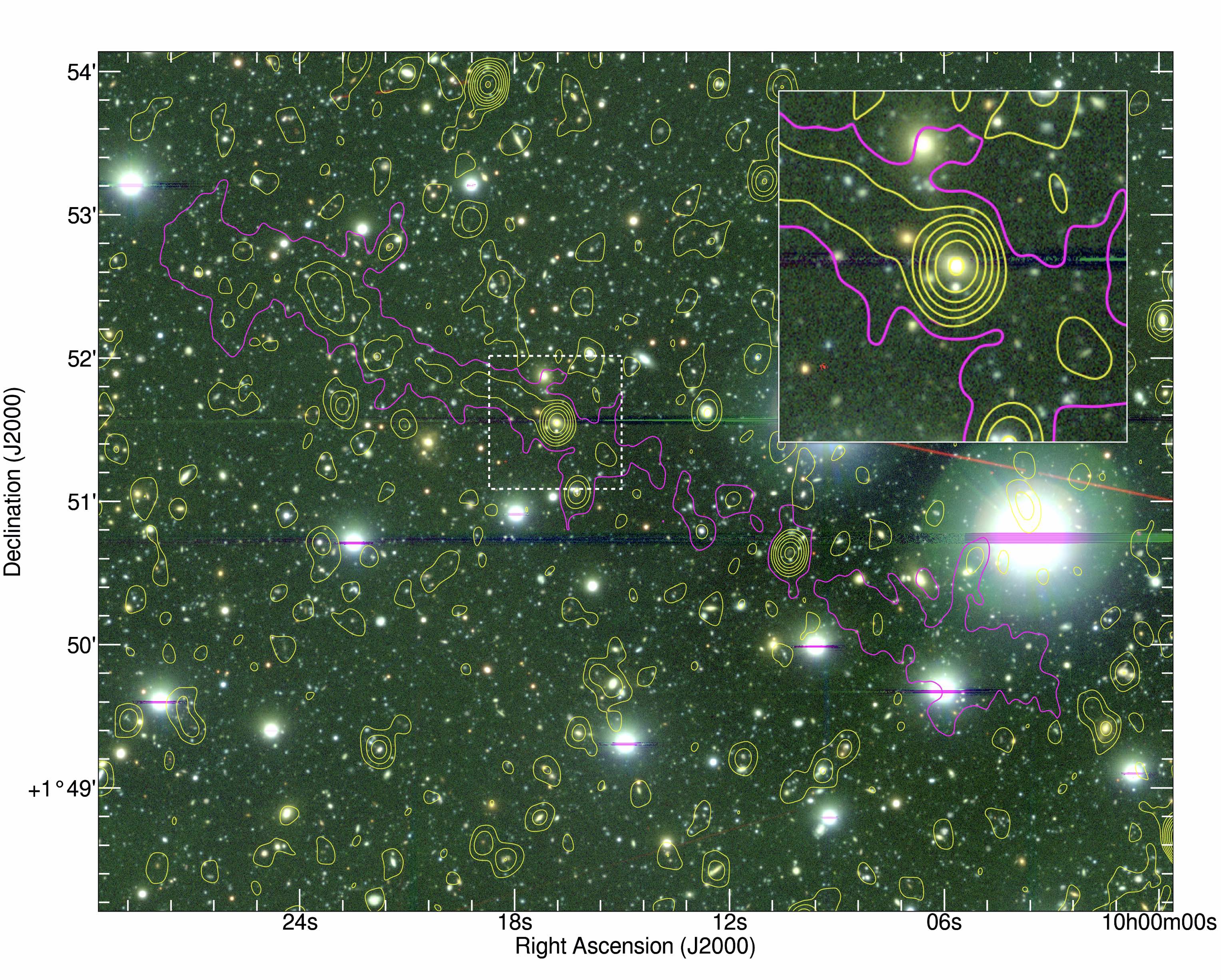}
\par\end{centering}
\caption{GRG2 as seen in MIGHTEE (yellow contours), overlaid on a composite optical image of HSC \emph{g}, \emph{r} and \emph{i} bands. Contours are shown at intervals of  $20\sqrt{3}^n\,\mu$Jy beam$^{-1}$, where $n=0,1,2...20$. The lowest contour level is $5\,\mu$Jy beam$^{-1}$ and is shown in magenta, as in Figure \ref{fig:grgs-Lband}, to highlight the full extent of the GRG.  The inset shows an enlargement of the core area.
\label{fig:grg2-optical}}
\end{figure*}

\begin{figure*}
\includegraphics[scale=0.15]{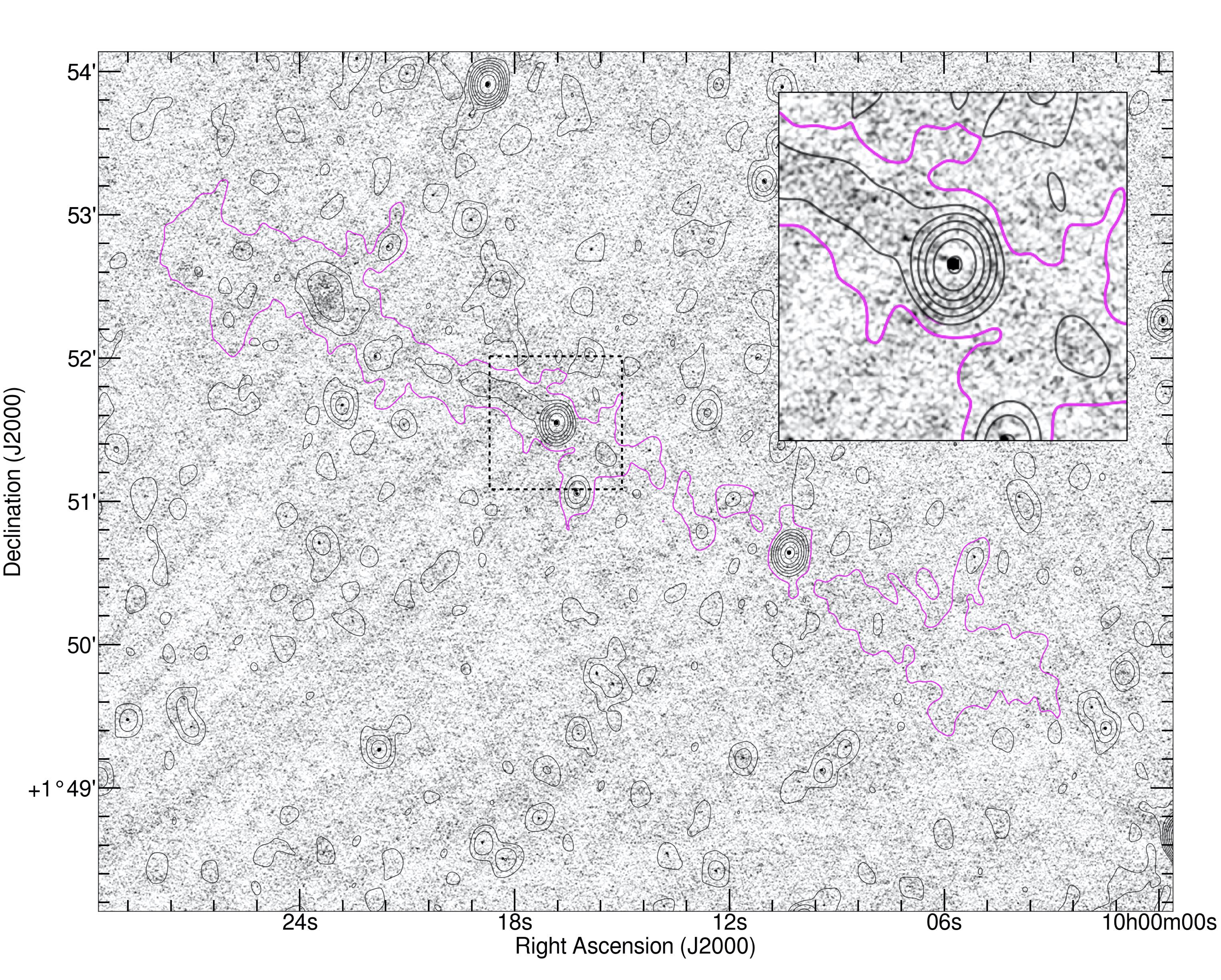}

\caption{GRG2 as seen in the VLA-3$\,$GHz data (background map) and in the MIGHTEE data (black contours). Contour levels as in Figure \ref{fig:grg1-optical}. At 3\,GHz a compact core is seen and diffuse emission in the northern hot-spot is faintly visible. \label{fig:grg2-radio}}
\end{figure*}

\subsubsection{MIGHTEE data}
GRG2 is shown in Figures \ref{fig:grgs-Lband}, \ref{fig:grg2-optical} and \ref{fig:grg2-radio}. The core is situated
at R.A.=10h00m16.84s and Dec=+01d51m33.0s. The associated host galaxy has a spectroscopic
redshift of $z=0.3363$ according to the zCOSMOS Bright catalogue (\citealp{lilly07}, \citealp{lilly09}).

Both lobes and a potential hot-spot within the northern lobe are clearly detected for
the first time in MIGHTEE, as is the entire northern jet and some weak emission from the southern jet. Based on these data, GRG2 has a projected angular size of 6.8\,arcmin and therefore a projected linear size of 2.04\,Mpc. Some
elongated emission is seen in MIGHTEE between the core and the southern
lobe, which may be attributable to the jet in that direction.

The unequal strength of detection of the two jets may imply that they are relativistic on these scales and lie away from the plane of the sky, so that their apparent surface brightness is being affected by relativistic beaming \citep{blanford79}.

Flux densities have been measured within the region outlined by the magenta contour shown in Figure \ref{fig:grgs-Lband}.

\subsubsection{VLA-3GHz data}
As seen in Figure \ref{fig:grg2-radio}, there is a hint of a VLA-3\,GHz detection of the northern lobe/hot-spot, though it does not appear in the source catalogue of
\citet{smolcic17a} due to its low SNR in these data. No hot-spot is seen in the southern lobe in MIGHTEE and this lobe is entirely undetected in the VLA-3\,GHz data. 

The core is detected in the VLA-3\,GHz observations and is reported
in the source catalogue of \citet{smolcic17b} with a 3\,GHz flux density
of $0.878\pm0.044$\,mJy. There is no evidence of any extended structure or unusual morphology of the core in these VLA data (see Figure \ref{fig:grg2-radio} inset).

\subsubsection{Radio power}
The MIGHTEE peak brightness of the GRG2 core ($0.72\pm0.01$\,mJy\,beam$^{-1}$) and the peak brightness from the smoothed VLA-3\,GHz map ($2.43\pm0.03$\,mJy\,beam$^{-1}$) are used to find a spectral slope of  $\alpha_{3{\rm GHz}}^{1.256{\rm GHz}}\sim0.29\pm0.04$. This is relatively flat, although slightly positive and is again consistent with typical cores of radio galaxies. Considering this and again assuming a spectral index of -0.8 for the lobes, the 1.2\,GHz radio power of GRG2 is  $\sim10^{24.1}$\,W\,Hz$^{-1}$. This is lower than most known GRGs \citep{dabhade20b} and slightly lower than that of GRG1.

\subsubsection{Host galaxy and AGN characterisation\label{host-grg2}}
The host galaxy appears elliptical in HSC imaging (see Figure \ref{fig:grg2-optical} inset). Despite having a zCOSMOS optical spectrum of relatively poor quality (bottom panel of Figure \ref{fig:spectrum}), it is typical of the absorption spectrum of a red, dead elliptical galaxy, with no emission features present.

GRG2 displays no signatures of AGN activity in the X-ray or mid-IR and
has no significant detection in the super-deblended {\em Herschel} photometry
of \citet{jin18}, with a combined SNR of all {\em Herschel} bands of only
$\sim0.8$. It is not identified as an AGN via SED
fitting (see Figure \ref{fig:SED-of-GRG}). Note that the SED fit is not improved by the inclusion of an AGN template despite the improvement at 24$\mu$m, since the outcome of the Fisher-test is guided much more by the worsening of the fit at IRAC-8$\mu$m. If the fit with AGN was preferred, the predicted AGN bolometric luminosity would correspond to an expected $L_X\sim 10^{42.6}$ erg\,s$^{-1}$, which is about 10 times higher than the formal X-ray limit at [2--10] keV \citep{civano16}. Since GRG2 is not X-ray detected, this check further supports the relatively low significance of the AGN component obtained from the Fisher-test.

The host galaxy of GRG2 has a stellar mass of $M_{*}=10^{10.8}$\,M$_{\odot}$ and a SFR
of $\sim0.5$\,M$_{\odot}$\,yr$^{-1}$, placing it well below the
main sequence of star formation. Its quiescent nature is further supported
by its classification as `red' based on its (\emph{NUV-r}) and (\emph{r-J}) colours
in COSMOS2015. 

We therefore conclude that GRG2, like GRG1, is a radiatively-inefficient LERG with
a massive, red, passive elliptical host. This is in line with expectations since LERGs tend to be hosted by galaxies which have redder colours, larger stellar masses and lower SFRs than HERG hosts (e.g. \citealt{best12,hardcastle13}). Little information currently exists about the SFRs of GRGs specifically, however \cite{clarke17} and \cite{dabhade20c} have found evidence for moderate SFRs in several individual GRG hosts. 

GRG2 resides in a smaller, less rich group than GRG1. The GRG2 host is the most massive of the five galaxies in the group (group ID 753; \citealt{knobel12}). 

We note that the compact source to the southwest of centre (labelled in Figure \ref{fig:grgs-Lband}) was also investigated as a potential position of the GRG core. However, it is not situated close to the
equidistant centre of the lobes and the associated optical counterpart
has a photometric redshift of $z=2.015$ in COSMOS15, so is less likely
to be the GRG host. It is considered an unrelated compact source and its contribution to the total flux density of the southern lobe has been removed.

\section{Discussion}
\subsection{Statistics and GRG sky density \label{sec:stats}}
\begin{figure*}
\includegraphics[scale=0.9]{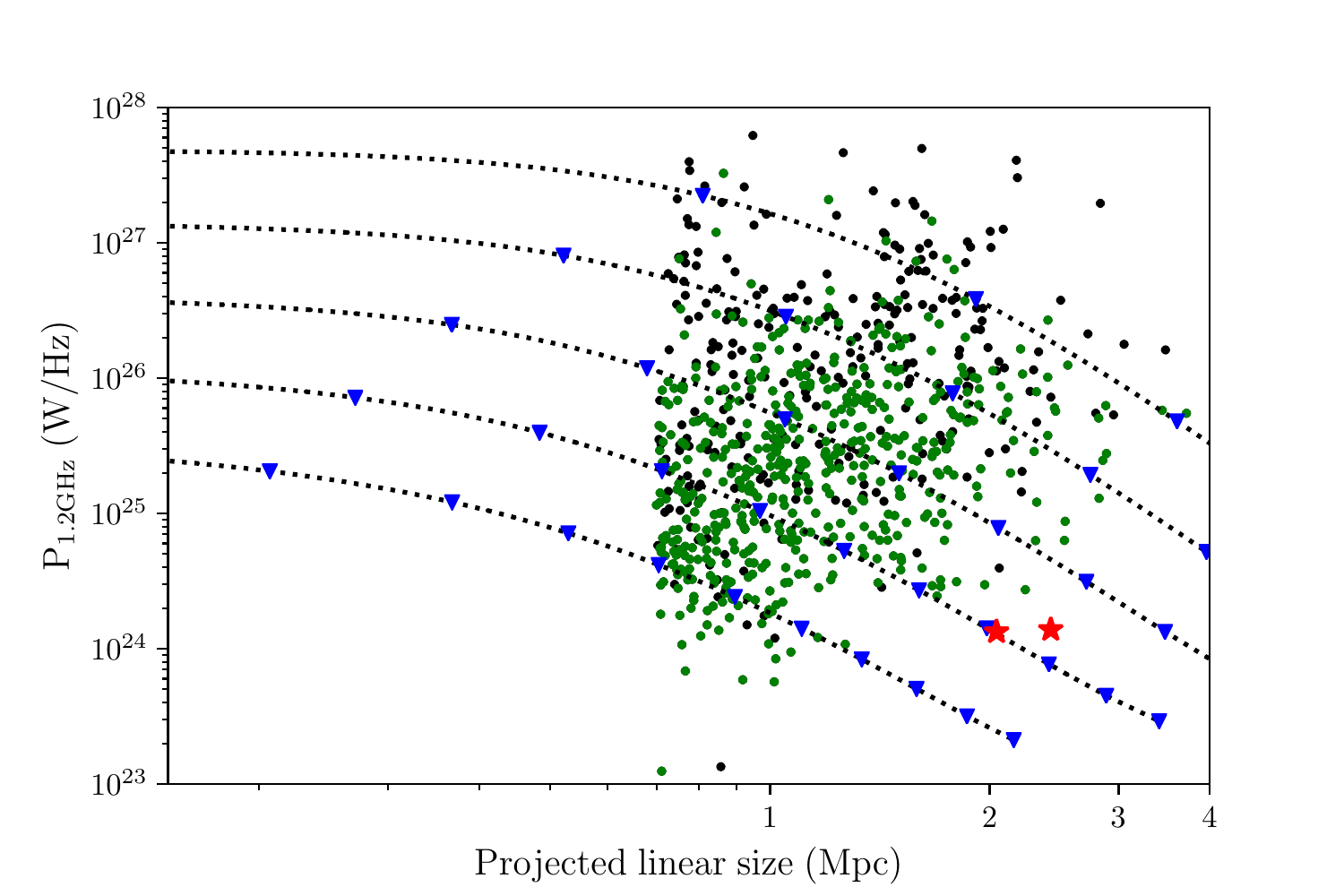}
\caption{The position in the 1.2\,GHz power-size (P-D) diagram of almost all known GRGs. Objects in the compilation catalogue of \protect\cite{dabhade20b} are shown as green ($z<0.5$) and black ($z\geq 0.5$) points. GRG1 and GRG2 are shown as red stars and are located in a previously unoccupied part of this diagram. For comparison, dotted lines show evolutionary tracks as determined by \protect\cite{hardcastle19} for radio galaxies with jet powers of (bottom to top) $10^{37}$, $10^{37.5}$, $10^{38}$, $10^{38.5}$ and $10^{39}$ W. The tracks show a lifetime of up to 1\,Gyr with blue triangles marking each hundred Myr. GRG1 and GRG2 are consistent with the evolution of a $\sim10^{37.5}$\,W jet after $\sim700$\,Myr.\label{fig:PD}}
\end{figure*}

\citet{dabhade20b} find that GRGs with sizes $>2$\,Mpc are extremely rare and comprise only 9\,percent of the GRG population at $z<1$. We now estimate the probability of finding two $>2$\,Mpc GRGs in such a small (1\,deg$^{2}$) sky area. For this purpose, our selection criteria can be considered to be GRGs with projected linear sizes $>2$\,Mpc out to a redshift of $z=0.5$ (and therefore with angular sizes of $>5.3$\,arcmin).  

We compare with the GRG sample of \citet{dabhade20}. They provide the most extensive survey of GRGs to-date, and therefore the most robust sky density estimate presently available. \citet{dabhade20} found 239 GRGs over a 424\,deg$^{2}$ sky area using the LoTSS survey. Only 7 of these have sizes larger than $2$\,Mpc, of which only one has $z<0.5$. This gives a sky density of 0.0023 GRGs per deg$^{2}$ for such objects.

We firstly assume that this sky density is the "true" GRG sky density of the Universe. We also assume independence of repeated events. That is,
that the presence of one GRG in the sky area does not affect whether
or not another is present, which is likely given the very different redshifts of the two GRGs presented here. We can thus employ Poisson statistics to estimate
the probability ($P$) of finding $k$ number of GRGs in a 1\,deg$^{2}$
sky area:

\begin{equation}
P(X=k)=\frac{\lambda^{k}e^{-\lambda}}{k!},
\end{equation}

where $\lambda$ is the expected number of occurrences and is equal to 0.0023 (which is the GRG sky density expected from the results of \citealt{dabhade20}, as above.) For the case of $k=2$, we find $P(X\ge2)=2.7\times10^{-6}$. Therefore, there is a probability of $2.7\times10^{-6}$ of finding two or more GRGs with projected linear sizes $>2$\,Mpc at $z<0.5$ within a 1\,deg$^{2}$ field, given our initial assumptions.

If the observed sky region is not strongly affected by cosmic (sample) variance
or other significant selection effects, the small p-value we find implies that either we have been exceptionally lucky in finding these two objects, or that our initial assumption does not hold and the true GRG sky density is significantly higher than previously known.

The most likely explanation for an under-estimated GRG sky density is the limited sensitivity to extended, diffuse emission
of GRG lobes in past surveys.  For example, LoTSS is incomplete to low-luminosity giants due to surface brightness sensitivity limitations, as illustrated in Figure 8 of \citet{hardcastle19}. While the components of GRG1 and GRG2 would exceed the 5$\sigma$ sensitivity limit of LoTSS, it would be difficult to identify these components as belonging to the same system if the low surface brightness emission joining them is not visible.

While the \textit{uv} coverage and angular resolution of LOFAR and MeerKAT are comparable, the point source sensitivity of MIGHTEE ($\sim11\,\mu$Jy at 144\,MHz, for $\alpha=-0.8$) is significantly deeper than LoTSS ($\sim71\,\mu$Jy at 144\,MHz). While it is non-trivial to directly compare surface brightness sensitivities of the two surveys, MIGHTEE is expected to be superior due to its depth. Low frequency surveys do benefit from being more sensitive to the steepest spectrum lobe emission of GRGs, however LoTSS would only surpass MIGHTEE in sensitivity where the spectral index is steeper than $\alpha\approx-1.7$.

A further explanation for a previously under-estimated GRG sky density is a dearth of highly sensitive, ancillary multi-wavelength observations. This would make it more difficult to determine whether or not lobes and hot-spots are associated with separate near-infrared/optical counterparts, particularly at higher redshifts. This issue is largely avoided by MIGHTEE-COSMOS thanks to the wealth of high-quality multi-wavelength data available in the field.

Therefore, GRGs may be yet more numerous than presently known, despite the rapid increase in the number of these giants discovered in recent years (e.g. \citealt{dabhade20,dabhade20b}).  This has also been suggested by other works, such as \citet{ishwara-chandra20} who discovered six new $\sim1$\,Mpc-sized GRGs out to $z\sim1.3$ in the ELAIS North 1 field with the GMRT. 

\subsection{GRG parameter space}

GRG1 and GRG2 do not seem atypical of large radio galaxies in that they are LERGs hosted
by ``red and dead'' elliptical galaxies, have flat spectrum cores
and exist in unremarkable environments (see \citealp{hardcastle20} for a review of radio AGN properties).

They do, however, have larger sizes and lower radio powers than most known GRGs. In fact, they lie in a heretofore unoccupied part of the GRG power-size (P-D) diagram. This is shown in Figure \ref{fig:PD}, in comparison to the 780 GRGs from the compilation catalogue of \cite{dabhade20b} for which $L$-band flux measurements are available from the NVSS or the literature (P. Dabhade, priv comm). For these, 1.2\,GHz radio powers have been calculated using the measured spectral index where available, or otherwise assuming a value of $\alpha=-0.8$.

Note that the lower limit (and likely more realistic) radio power of GRG1 ($P_{1.2}=10^{24.1}$ W\,Hz$^{-1}$; see Section \ref{GRG1-P}) has been used in Figure \ref{fig:PD}. However, GRG1 would still sit slightly below all other GRG of the same size on this diagram if the conservative upper limit to its radio power was used.

It is therefore possible that these new MIGHTEE observations are uncovering an unexplored realm of GRG parameter space, hitherto invisible to previous surveys due to the limitations discussed in Section \ref{sec:stats}. These may be the low-luminosity giants predicted by the P-D tracks of, for example, \cite{kaiser97}, \citet{shabala08} and \citet{hardcastle19}. 

To illustrate this, Figure \ref{fig:PD} also shows the evolutionary tracks of radio galaxies of various jet powers, as determined by \cite{hardcastle19}. These are shown for an environment of halo mass $10^{13}$\,M$_\odot$, observing frequency 1.2\,GHz and $z=0.25$. GRG1 and GRG2 are consistent with the evolution of a $\sim10^{37.5}$\,W jet after $\sim700$\,Myr, under these conditions.

\subsection{Implications}

While our analysis has considered only enormous ($>2$\,Mpc) objects, if radio galaxies must grow to reach this size, then we may expect to similarly uncover in our data previously undetected GRGs with smaller sizes. The full 20\,deg$^{2}$ MIGHTEE survey will provide an excellent resource for such studies when it is completed in the near future. Based on the GRG sky density we observe in COSMOS (2 per deg$^{2}$), we could uncover $\sim40$ GRGs with the full survey. With MeerKAT simultaneously facilitating excellent sensitivity, sky coverage and $uv$ coverage with relatively high resolution, and the upcoming Square Kilometre Array (SKA) providing even better angular resolution, we may expect to discover many more diffuse, extended GRGs in the near- to mid-future.

The existence of a larger population of GRGs may have
implications for our understanding of the AGN duty cycle. Low frequency follow-up observations of these objects can provide spatially-resolved spectral information, which will help us to characterise the intermittency of the radio AGN activity. Such information is crucial for our understanding of whether AGN feedback across subsequent accretion episodes remains constant or decreases with time, enabling predictions of the overall life-cycle of AGN activity relative to the host. This is essential for advancing our current understanding of the extent to which AGN-driven mechanical feedback can alter the star-forming content of the host galaxy \citep{croton06}.

\section{Conclusions}
Thanks to the impressive capabilities of the MeerKAT telescope, two new
giant radio galaxies have been identified in a 1\,deg$^{2}$ MIGHTEE
survey of the COSMOS field. Both are "red and dead" LERGs and are among the largest known GRGs with
sizes of $>2$\,Mpc. However, they have low radio powers which place them in a previously unpopulated part of GRG parameter space. Due to the diffuse nature of the jets and lobes, these
objects were resolved out and undetected in previous surveys of the
COSMOS field, including sensitive 3\,GHz observations with the VLA. 

The probability of finding at least two such GRGs in a small, 1\,deg$^2$ field is only
$2.7\times10^{-6}$, based on wide-field observations with LOFAR. Therefore, our findings provide strong evidence that GRGs may be far more numerous than previously thought. It is only
with new radio surveys such as MIGHTEE, providing excellent extended
brightness sensitivity, that this `hidden' population of GRGs can be revealed. Systematic searches across the full MIGHTEE survey are expected to yield detections of many more GRGs and low frequency follow-up can help reveal important information about the AGN duty cycle. These are tantalizing hints of what the future SKA will ultimately uncover with its simultaneously excellent angular resolution and surface brightness sensitivity.

\section*{Acknowledgements}

We thank the anonymous referee for providing useful comments which have strengthened the paper. We thank Pratik Dabhade for helpful discussions and for generously providing their GRG catalogue. The MeerKAT telescope is operated by the South African Radio Astronomy Observatory (SARAO; www.ska.ac.za), which is a facility of the National Research Foundation (NRF), an agency of the Department of Science and Innovation. JD, SVW, IH and MP acknowledge the financial assistance of SARAO. MP, SMR and LL acknowledge the financial assistance of the NRF. Opinions expressed and conclusions arrived at, are those of the authors and are not necessarily to be attributed to the NRF. 

MJJ and IH acknowledge support from the UK Science and Technology Facilities Council [ST/N000919/1]. MJJ, IH and IHW acknowledge support from the Oxford Hintze Centre for Astrophysical Surveys which is funded through generous support from the Hintze Family Charitable Foundation. IH thanks the Rhodes University Centre for Radio Astronomy Techniques and Technologies (RATT) for the provision of computing facilities. ID is supported by the European Union's Horizon 2020 research and innovation program under the Marie Sk\l{}odowska-Curie grant agreement No 788679. MP, JC, FXA, MV, KT and MG acknowledge financial support from the Inter-University Institute for Data Intensive Astronomy (IDIA). The research of OS is supported by the South African Research Chairs Initiative of the NRF. YA acknowledges support by NSFC grant 11933011. RB acknowledges support from the Glasstone Foundation. LM, MV and IP acknowledge financial support from the Italian Ministry of Foreign Affairs and International Cooperation (MAECI Grant Number ZA18GR02) and the South African NRF (Grant Number 113121) as part of the ISARP RADIOSKY2020 Joint Research Scheme. JA acknowledges financial support from the Science and Technology Foundation (FCT, Portugal) through research grants PTDC/FIS-AST/29245/2017, UID/FIS/04434/2019, UIDB/04434/2020 and UIDP/04434/2020. NA acknowledges funding from the Science and Technology Facilities Council (STFC) Grant Code ST/R505006/1. ZR acknowledges financial support from the South African Astronomical Observatory which is a facility of the South African NRF. NM acknowledges support from the Bundesministerium f{\"u}r Bildung und Forschung (BMBF) D-MeerKAT award 05A2017. KT acknowledges support from the Inter-University Institute for Data Intensive Astronomy (IDIA). MB acknowledges support from the ERC-Stg DRANOEL, no 714245.

We acknowledge the use of the Ilifu cloud computing facility - www.ilifu.ac.za, a partnership between the University of Cape Town, the University of the Western Cape, the University of Stellenbosch, Sol Plaatje University, the Cape Peninsula University of Technology and the South African Radio Astronomy Observatory.  The Ilifu facility is supported by contributions from the Inter-University Institute for Data Intensive Astronomy (IDIA - a partnership between the University of Cape Town, the University of Pretoria and the University of the Western Cape), the Computational Biology division at UCT and the Data Intensive Research Initiative of South Africa (DIRISA). This work made use of the CARTA (Cube Analysis and Rendering Tool for Astronomy) software (DOI 10.5281/zenodo.3377984 - https://cartavis.github.io).

The Hyper Suprime-Cam (HSC) collaboration includes the astronomical communities of Japan and Taiwan, and Princeton University. The HSC instrumentation and software were developed by the National Astronomical Observatory of Japan (NAOJ), the Kavli Institute for the Physics and Mathematics of the Universe (Kavli IPMU), the University of Tokyo, the High Energy Accelerator Research Organization (KEK), the Academia Sinica Institute for Astronomy and Astrophysics in Taiwan (ASIAA), and Princeton University. Funding was contributed by the FIRST program from Japanese Cabinet Office, the Ministry of Education, Culture, Sports, Science and Technology (MEXT), the Japan Society for the Promotion of Science (JSPS), Japan Science and Technology Agency (JST), the Toray Science Foundation, NAOJ, Kavli IPMU, KEK, ASIAA, and Princeton University.

\section{Data availability}
The data underlying this article were accessed from the South African Radio Astronomy Observatory (SARAO; www.ska.ac.za). The derived data generated in this research will be shared on reasonable request to the corresponding author.

\bibliographystyle{mn2e}
\bibliography{mightee_grg}

\subsubsection*{Author Affiliations}
$^{\uct}$Department of Astronomy, University of Cape Town, Private Bag X3, Rondebosch 7701, South Africa
\\$^{\ox}$Astrophysics, Department of Physics, University of Oxford, Keble Road, Oxford OX1 3RH, UK
\\$^{\rhod}$Department of Physics and Electronics, Rhodes University, PO Box 94, Makhanda, 6140, South Africa
\\$^{\sarao}$South African Radio Astronomy Observatory, 2 Fir Street, Black River Park, Observatory, Cape Town 7925, South Africa
\\$^{\idiauwc}$Inter-University Institute for Data Intensive Astronomy, and Department of Physics and Astronomy, \\University of the Western Cape, Robert Sobukwe Road, 7535 Bellville, Cape Town, South Africa 
\\$^{\uwc}$Department of Physics and Astronomy, University of the Western Cape, Robert Sobukwe Road, 7535 Bellville, Cape Town,\\ South Africa
\\$^{\cea}$CEA, IRFU, DAp, AIM, Universit\'e Paris-Saclay, Universit\'e de Paris, CNRS, F-91191 Gif-sur-Yvette,\\ France 
\\$^{\inafm}$INAF - Osservatorio Astronomico di Brera, via Brera 28, I-20121, Milano, Italy \& via Bianchi 46, I-23807, Merate, Italy
\\$^{\herts}$Centre for Astrophysics Research, Department of Physics, Astronomy and Mathematics, University of Hertfordshire, College Lane, \\ Hatfield AL10 9AB, UK 
\\$^{\csiro}$CSIRO Astronomy and Space Science, PO Box 1130, Bentley WA 6102, Australia 
\\$^{\jaA}$Instituto de Astrof\'{i}sica e Ci\^{e}ncias do Espa\c co, Universidade de Lisboa, OAL, Tapada da Ajuda, PT1349-018 Lisboa, Portugal
\\$^{\jaB}$Departamento de F\'{i}sica, Faculdade de Ci\^{e}ncias, Universidade de Lisboa, Edif\'{i}cio C8, Campo Grande
\\$^{\yaA}$Purple Mountain Observatory and Key Laboratory for Radio Astronomy, Chinese Academy of Sciences, Nanjing, China
\\$^{\yaB}$School of Astronomy and Space Science, University of Science and Technology of China, Hefei, Anhui, China 
\\$^{\brienz}$ Dipartimento di Fisica e Astronomia, Universit\'{a} di Bologna, via P. Gobetti 93/2, 40129, Bologna, Italy
\\$^{\inafb}$INAF, Istituto di Radioastronomia, Via Gobetti 101, 40129, Bologna, Italy 
\\$^{\mbA}$Hamburger Sternwarte, University of Hamburg, Gojenbergsweg 112, 21029 Hamburg, Germany 
\\$^{\idiauct}$The Inter-University Institute for Data Intensive Astronomy (IDIA), Department of Astronomy, University of Cape Town,\\ Private Bag X3, Rondebosch, 7701, South Africa
\\$^{\syd}$School of Science, Western Sydney University, Locked Bag 1797, Penrith, NSW 2751, Australia 
\\$^{\nmA}$Faculty of Physics, Ludwig-Maximilians-Universit\"at, Scheinerstr. 1, 81679 Munich, Germany 
\\$^{\durh}$Centre for Extragalactic Astronomy, Department of Physics, Durham University, Durham, DH1 3LE, UK 
\\$^{\saao}$South African Astronomical Observatory, P.O. Box 9, Observatory 7935, Cape Town, South Africa
\\$^{\nrao}$National Radio Astronomy Observatory, 1003 Lopezville Road, Socorro, NM 87801, USA
\\$^{\smrA}$A\&A, Department of Physics, Faculty of Sciences, University of Antananarivo, B.P. 906, Antananarivo 101, Madagascar 
\\$^{\pret}$Department of Physics, University of Pretoria, Private Bag X20, Hatfield 0028, South Africa 
\\$^{\icrar}$International Centre for Radio Astronomy Research, Curtin University, Bentley, WA 6102, Australia 
\\$^{\GEPI}$ GEPI, Observatoire de Paris, CNRS, Universit\'e Paris Diderot, 5 place Jules Janssen, 92190 Meudon, France
\\$^{\RATSS}$ Centre for Radio Astronomy Techniques and Technologies, Department of
Physics and Electronics, Rhodes University,\\ Grahamstown 6140, South Africa
\\$^{\USN}$ USN, Observatoire de Paris, CNRS, PSL, UO, Nan\c cay, France

\label{lastpage}
\end{document}